\documentclass{Interspeech}

\interspeechcameraready 
\usepackage{cite}
\usepackage{amsmath,amssymb,amsfonts, pifont,graphicx,amssymb,booktabs,hyperref,multirow,booktabs,caption,color,cite,amssymb, xcolor}
\usepackage{algorithmic}
\usepackage{graphicx}
\usepackage{textcomp}
\usepackage{xcolor}
\usepackage{marvosym}
\usepackage{microtype}
\usepackage{subfigure}
\usepackage{adjustbox}
\usepackage{etoolbox}

\title{FreeCodec: A disentangled neural speech codec with fewer tokens}

\author[affiliation={1,*}]{Youqiang}{Zheng}
\author[affiliation={1,\dagger}]{Weiping}{Tu}
\author[affiliation={2}]{Yueteng}{Kang}
\author[affiliation={2}]{Jie}{Chen}
\author[affiliation={2}]{Yike}{Zhang}
\author[affiliation={1}]{Li}{Xiao}
\author[affiliation={1}]{Yuhong}{Yang}
\author[affiliation={2}]{Long}{Ma}

\affiliation{NERCMS, School of Computer Science}{Wuhan University}{China}
\affiliation{}{Tencent Youtu Lab}{China}
\email{youqiangzheng@whu.edu.cn, tuweiping@whu.edu.cn, yuetengkang@tencent.com}

\usepackage{comment}

\begin{document}

\maketitle
\renewcommand{\thefootnote}{\fnsymbol{footnote}}
\footnotetext[1]{Work is done during the internship at Tencent YouTu Lab}
\renewcommand{\thefootnote}{\arabic{footnote}}
\renewcommand{\thefootnote}{\fnsymbol{footnote}}
\footnotetext[2]{Corresponding author}
\renewcommand{\thefootnote}{\arabic{footnote}}
\begin{abstract}
    
    Neural speech codec is a crucial component in generative tasks such as speech resynthesis and zero-shot TTS. However, most works exhibit degraded performance with fewer tokens due to low coding efficiency in modeling complex coupled information. In this paper, we propose a self-supervised disentangled neural speech codec named FreeCodec. It employs distinct frame-level encoders to decompose intrinsic speech properties into separate components and adopts enhanced decoders to reconstruct speech signals. By encoding and quantizing the different frame-level information with dedicated quantizers, FreeCodec gets higher encoding efficiency with 57 tokens. Furthermore, our proposed method can be applied flexibly in reconstruction and disentanglement scenarios with different training strategies.
Subjective and objective experimental results demonstrate that our framework outperforms existing methods in both reconstruction and disentanglement tasks. 
\end{abstract}

\section{Introduction}

\label{sec:intro}
Neural speech codecs are widely used to compress speech signals for a limited number of bits with minimal distortion. 
Compared to traditional parametric algorithms \cite{rowe2011codec,supplee1997melp}, it has progressed significantly in medium- or low-bitrate scenarios. 
With the development of large language models (LLM), the discrete codes of neural speech codecs play a pivotal role in LLM-driven generative speech models~\cite{wang2023neural, borsos2023audiolm, defossez2024moshi}. In general, the fewer tokens, the lower the bitrates while remaining high-quality, which is the goal of neural speech codecs.

The existing mainstream neural speech codecs \cite{zeghidour2022soundstream, defossez2022high,jiang2023latent, zheng2024supercodec, ji2024wavtokenizer} rely on the architecture of VQ-VAE \cite{van2017neural}. An encoder, vector quantization layers, and a decoder are learned in end-to-end (E2E) by data-driven. 
These techniques utilize vector quantization layers to discrete the continuous latent features from the encoder.
Recently, many studies have explored disentanglement methods to enhance the quality of reconstructed speech. 
There are two mainstream methods for disentanglement: 1) The supervised method, and 2) The unsupervised method. 
The supervised method, e.g. FACodec in NaturalSpeech3\cite{ju2024naturalspeech}, considers disentanglement by the amount of data annotation such as Phone, F0, Speaker labels, etc. Although it is efficient in disentanglement scenarios, it operates at higher bitrates.

The unsupervised method~\cite{polyak2021speech, ren2024fewer,zhang2024speechtokenizer,li2024single, liu2024semanticodec, guo2024fireredtts, guo2024lscodec} is an implicit disentangled technique that usually focuses on using disentanglement to enhance coding efficiency.
On the one hand, TiCodec \cite{ren2024fewer} and SingleCodec \cite{li2024single} incorporate an additional global encoder to extract time-invariant information from speech. The quantization of the extracted global embedding is unnecessary in this way. These methods reduce the redundancy of frame-level information to attain improved encoding efficiency and exhibit improved performance using one or two codebooks at reconstruction scenarios.
On the other hand, in \cite{zhang2024speechtokenizer, liu2024semanticodec}, self-supervised learning models are employed to factorize semantic and acoustic representations within vector quantization layers, achieving more effective compression compared to conventional neural speech codecs such as EnCodec\cite{defossez2022high} and Descript-audio-codec (DAC)\cite{kumar2024high}. 

The aforementioned methods have demonstrated a strong ability to improve the quality of reconstructed speech using disentanglement techniques. However, these methods
has failed to get a balance between reconstruction and disentanglement. Moreover, speech includes several attributes(not just global and non-global), and each of them should be modeled using a module \cite{jiang2023mega}. Inspired by this, we explore a self-supervised disentanglement of representations framework, which can be used flexibly in reconstruction and disentanglement scenarios.

\begin{figure*}[h]
    \hspace*{1.cm}\subfigure[The architecture of FreeCodec]{\includegraphics[width=0.54\textwidth, trim={0 0 0cm 0}, clip]{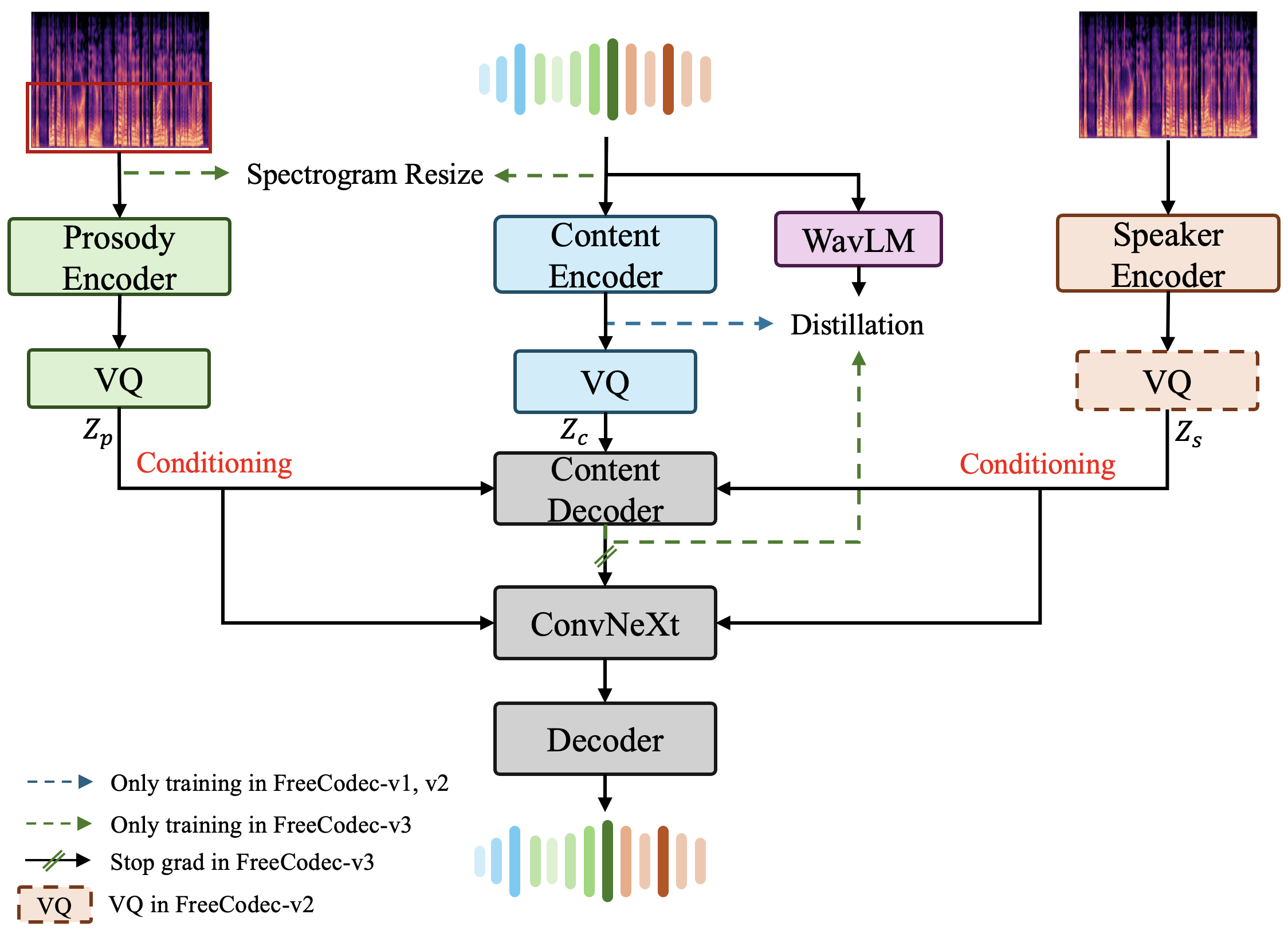}
     \label{a}}
    \hspace*{1.2cm}\subfigure[Prosody Encoder]{\includegraphics[width=0.13\textwidth, trim={0 0 1cm 0}, clip]{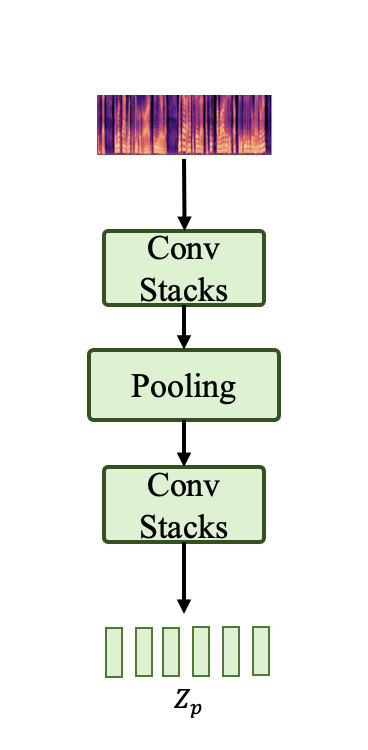}
    \label{b}}
    \caption{The architecture of FreeCodec and prosody encoder. In Fig.\ref{a}, we show different versions of the FreeCodec training paradigm in different scenarios. In Fig.\ref{b}, the prosody encoder removes the speaker information by a max pooling layer. }
    \label{fig:architecture}
    \vspace{-0.5cm}
\end{figure*}

 In this paper, we propose a self-supervised disentangled neural speech codec - FreeCodec. It models complex speech into intrinsic attributes(speaker, prosody, and content) in the encoder and disentangles speaker information explicitly. 
We adopt different frame-level representations for different attributes, enabling more effective quantization and higher compression. In addition, we adopt an improved 
decoder to improve information reconstruction. Our main contributions are as follows:

\begin{itemize}
    \item We propose FreeCodec, a self-supervised disentangled neural speech codec that encodes intrinsic properties in speech to disentangle speaker information explicitly and adopts enhanced decoders to get better reconstruction.
    \item We show that our proposed framework can be flexibly used in reconstruction(e.g., zero-shot TTS, speech compression) and disentanglement(e.g., voice conversion) scenarios when using different training strategies.
    \item Our proposed method using approximately 57 tokens per second, surpasses the existing state-of-the-art models in subjective and objective evaluation.
\end{itemize}

\section{Proposed Method}

\subsection{Overall}
\vspace{-0.2cm}
As illustrated in Fig.\ref{a}, our proposed method consists of three components: encoders, quantizers, and decoders.
Unlike existing works, our encoder proposes a more detailed modeling focus on different intrinsic properties in human speech. We introduce three types of encoders to encode content, speaker, and prosody (in addition to the content and the speaker) information, respectively. Then the quantization layers produce compressed representations. 
Finally, the improved decoders, consisting of a content decoder, a backbone module, and an up-sampling decoder, reconstruct the speech signal from the compressed latent representations. 
In addition, we utilize different training strategies to provide three versions for reconstruction and disentanglement scenarios. 
The details of training strategies are described in Section \ref{quantization} and \ref{decoder}. 
\vspace{-0.3cm}

\subsection{Encoders}
\textbf{Speaker Encoder} Existing approaches assume that a global embedding can represent time-invariant information, such as the characteristics of the speaker and speaking style.
Here, we follow this unsupervised manner and further extract the speaker's information more precisely.
We utilize a pre-trained speaker encoder, ECAPA-TDNN \cite{desplanques2020ecapa}, a state-of-the-art speaker recognition network based on convolution neural networks with an attentive statistics pooling layer.
A mel-spectrogram sampled from the raw speech signal is fed into the speaker encoder to get one global timbre vector. 

\textbf{Content Encoder} The architecture of the content encoder follows SuperCodec \cite{zheng2024supercodec} encoder using (2, 4, 5, 8) as strides, a number \textbf{${B}_{enc}$} = 4 of convolution blocks.
It indicates a total down-sampling of 320 times and outputs 256-dimensional content features with a frame rate of 50 Hz from 16 kHz speech. 
In order to reduce the redundancy of the content encoder, we use a self-supervised model to explicitly model the content information, as shown in Fig.\ref{a}.

\textbf{Prosody Encoder} The prosody encoder extracts the information apart from the speaker and content information, as shown in Fig.\ref{b}.
In \cite{ren2022prosospeech, jiang2024megatts}, the first 20 bins in each mel-spectrogram frame are taken as input to extract prosody because it contains almost complete prosody and much less speaker and content information than the full band.
Following the related work \cite{jiang2024megatts}, we adopt the prosody encoder consisting of two convolution stacks, a max pooling layer with a stride of 8 to remove content and speaker information further.
Our proposed method sets the FFT and hop size to 1024 and 320. With these setups, the prosody encoder results in roughly a frame rate of 7 Hz feature embeddings with 256 dimensions.
\vspace{-0.4cm}
\subsection{Quantization}
\label{quantization}
We adopt different methods to quantize different features. 
For the content and prosody information, we adopt a plain vector quantizer with one codebook, and the codebook size is set to 256.
As for the speaker embedding, we use two types: continuous representation for FreeCodec-v1 and FreeCodec-v3 and discrete representation for FreeCodec-v2.
Specifically, we compress the speaker embedding by group vector quantization (GVQ) in FreeCodec-v2 for speech coding. It divides the speaker embedding into eight groups that are quantized by one codebook with 1024 codebook size respectively.
As for FreeCodec-v1 and FreeCodec-v3, we provide the continuous representation to the decoder for better reconstruction in such as zero-shot TTS and voice conversion scenarios, similar to \cite{ju2024naturalspeech, pan2024promptcodec}.
\vspace{-0.3cm}
\subsection{Improved Decoders}

FreeCodec does not merely rely on a mirrored upsampling decoder.  Prior to upsampling, we first employ a content decoder consisting of a 4-layer Transformer encoder to enhance semantic modeling.  Then, we use ConvNeXt~\cite{liu2022convnet} as a fundamental backbone to condition the prosody and speaker representations furtherly. Finally, a mirrored decoder upsampling structure is employed to reconstruct speech signals. It uses (8, 5, 4, 2) as strides, resulting in a total upsampling of 320 times.  
\vspace{-0.3cm}

\subsection{Training Strategy}
\label{decoder}
We incorporate adversarial training to promote perceptual quality, using a multi-scale STFT-based (MS-STFT) discriminator. The training loss of the proposed method comprises five components: reconstruction loss $\lambda_{\text {rec}}$, VQ commitment loss $\lambda_{\text {vq }}$, content loss $\lambda_{\text {c }}$, feature matching loss $\lambda_{\text {feat }}$, and adversarial losses $\lambda_{\text {adv }}$. The reconstruction loss, feature loss, and adversarial losses follow  EnCodec \cite{defossez2022high}. We extract the last layer representation from a pre-trained WavLM-Large model~\cite{chen2022wavlm} as the semantic learning target and use cosine similarity loss as the content loss. 

In FreeCodec-v1 and FreeCodec-v2, we use it to reduce the redundancy of the content encoder. 
It maximizes the cosine similarity at the level of the dimensions across all timesteps between the outputs of the content encoder and semantic learning target.
In FreeCodec-v3, we only use the semantic learning target at the decoder to prevent additional speaker information from leaking to the content encoder and quantizer. We also utilize spectrogram-resize based
data augmentation on the prosody and content encoder in the training \cite{freevc}.
This approach achieves better performance in disentanglement scenarios.

Overall, the generator loss is defined as, 
\vspace{-0.2cm}
\begin{equation}
\mathcal{L}_G=\lambda_{\text {adv }} \cdot \mathcal{L}_{a d v}+\lambda_{\text {feat }} \cdot \mathcal{L}_{\text {feat }}+\lambda_{\text {rec }} \cdot \mathcal{L}_{\text {rec }}+\lambda_{vq} \cdot \mathcal{L}_{vq} + \lambda_{c} \cdot \mathcal{L}_{c},
\end{equation}
where these hyper-parameters $\lambda_{\text {adv }} = 3$, $\lambda_{\text {feat }} = 3$, $\lambda_{\text {rec }} = 1$, $\lambda_{\text {vq }} = 1$, and $\lambda_{\text {c }} = 10$.

\section{Experimental Setup}

\subsection{Training Details and Baselines}
We trained our model on LibriSpeech \cite{panayotov2015librispeech}, which consists of approximately 1000 hours of speech at 16kHz. We use train-clean-100, train-clean-360, and train-other-500 subsets for training. For a fair comparison, we adopt two recent neural codecs, TiCodec and Descript-audio-codec (DAC), which have demonstrated success in the domain of neural speech codecs.
The baselines are re-trained with 1 and 2 codebooks, indicating 0.5 kbps and 1 kbps at 16 kHz sampling rate. 
FreeCodec and TiCodec are trained on two V100 GPUs with 400 k iterations and a batchsize of 20 per GPU. 
DAC is trained on two V100 GPUs with 800 k iterations and a batchsize of 10 per GPU.
In addition, we also consider several open-source speech codecs as baselines, Speechtokenizer at 3 kbps, and FACodec without acoustic details at 2.4 kbps, and SemantiCode at 1.3 kbps, and Wavtokenizer-small at 0.9 kbps.
For Wavtokenizer-small, we use the 24 kHz pre-trained model to synthesize speech, corresponding to the same compression rate of 0.6 kbps for the 16 kHz sampling rate. 

As for the voice conversion, we include disentangled codecs TiCodec at 0.5 kbps and 1 kbps, and FACodec without detail tokens at 2.4 kbps. In addition, three baseline models are selected to be compared with FreeCodec-v3:
two text-based models—VQMIVC \cite{wang2021vqmivc} and YourTTS \cite{casanova2022yourtts}, a self-supervised learning model—Wav2vec-vc \cite{lim2024wav2vec}, which are trained on the VCTK datasets. 
\begin{table*}[h]
\centering
        \caption[short]{The objective evaluation of reconstruction quality based on VCTK and LibriSpeech Test-Clean corpus. \scalebox{0.7}{$\blacklozenge$} represents the reproduced model following the official implementation. \scalebox{0.7}{$\clubsuit$} means the results inferenced from the official checkpoints.
        \textbf{Bold} is the best result and \underline{underline} is the second-best result.
        }
        \vspace{-0.2cm}
        \setlength{\tabcolsep}{0.6mm}{%
        \begin{tabular}{lllccccc|cccc}
        \toprule
\multicolumn{1}{c}{\multirow{2}{*}{Model}}      & \multirow{2}{*}{Sampling rate} & \multirow{2}{*}{Bandwidth} & \multirow{2}{*}{Token/s} & \multicolumn{4}{c|}{VCTK}                                                                                                          & \multicolumn{4}{c}{Test-clean}                                                                                                    \\
\multicolumn{1}{c}{}                 &             &       &        & UTMOS$\uparrow$                 & STOI$\uparrow$                 & WARP-Q$\downarrow$             & SECS$\uparrow$                  & UTMOS$\uparrow$                & STOI$\uparrow$                 & WARP-Q$\downarrow$             & SECS$\uparrow$                  \\ \hline
Target                    & -               & -      & \multicolumn{1}{c}{-}    & \multicolumn{1}{c}{4.085}       & \multicolumn{1}{c}{-}          & \multicolumn{1}{c}{-}          & \multicolumn{1}{c|}{-}         & \multicolumn{1}{c}{4.086}      & \multicolumn{1}{c}{-}          & \multicolumn{1}{c}{-}          & \multicolumn{1}{c}{-}          \\
SpeechTokenizer$^{\scalebox{0.6}{$\clubsuit$}}$     & 16 kHz    & 3 kbps         & 300     &3.953                         & 0.868                          & 2.048                          &0.915                          & 3.848                       &\textbf{0.908}                          & \textbf{2.034}                          & \textbf{0.954}                          \\
FACodec$^{\scalebox{0.6}{$\clubsuit$}}$ & 16 kHz &2.4 kbps &240 &\textbf{4.085} &0.855 & 2.026 & \textbf{0.929} & 3.616 & 0.876 & 2.170 &  0.943 \\
SemantiCodec$^{\scalebox{0.6}{$\clubsuit$}}$     & 16 kHz        & 1.3 kbps       & 100     & 3.334      & 0.853                          & 2.078                          & 0.868                          & 2.922                        & 0.879                          & \underline{2.049}                          & 0.936                          \\

TiCodec$^{\scalebox{0.6}{$\blacklozenge$}}$  & 16 kHz    & 1 kbps         & 100       & 3.584       & 0.879  & 2.333                          & 0.856                          & 3.616                          & 0.881                          & 2.354                          & 0.908                          \\
DAC$^{\scalebox{0.6}{$\blacklozenge$}}$    & 16 kHz          & 1 kbps         & 100      & 3.780       & 0.904                          & 2.251                          & 0.883                          & 3.790                          & \underline{0.901}                 & 2.274                          & 0.920                          \\ 
WavTokenizer$^{\scalebox{0.6}{$\clubsuit$}}$     & 24 kHz     & 0.9 kbps       & 75        & 3.296      & 0.832                          & 2.192                          & 0.811                          & 3.792                         & 0.897                          & 2.135                          & 0.904                          \\
TiCodec$^{\scalebox{0.6}{$\blacklozenge$}}$  & 16 kHz    & 0.5 kbps       & 50      & 3.421       & 0.825     & 2.578                          & 0.797                          & 3.307                          & 0.821                          & 2.614                          & 0.855                          \\
DAC$^{\scalebox{0.6}{$\blacklozenge$}}$     & 16 kHz     & 0.5 kbps          & 50       & 3.476    & 0.852       & 2.550                          & 0.804                          & 3.543                          & 0.859                          & 2.504                          & 0.883                          \\ 

\hline
FreeCodec-v1              & 16 kHz    & 0.45 kbps                  & 57       &  \underline{4.034}   & \textbf{0.918}    & \textbf{1.966}                 &  \underline{0.919}                 & \textbf{4.085}                 & 0.892                          & 2.195 & 0.944 \\
w/o.${L}_{content}$           & 16 kHz         & 0.45 kbps          &57        & 3.805     & \underline{0.908} & \underline{1.994} & 0.893 & 3.631                          & 0.869                          & 2.308                          & 0.925                          \\
FreeCodec-v2                      & 16 kHz        & 0.45 kbps         &57        & 3.921 & 0.900      & 2.190                          & 0.846                          & \underline{3.984} & 0.893 & 2.230                          & 0.896                          \\
w/o.${L}_{content}$         & 16 kHz          & 0.45 kbps           &57       & 3.578            & 0.892                          & 2.175                          & 0.848                          & 3.571                          & 0.885                          & 2.223                          & 0.904                          \\ \bottomrule
\end{tabular}
\label{speech}
}
\end{table*}

\subsection{Evaluation}
We evaluate FreeCodec from two aspects: \textbf{1) Reconstruction Quality}. We conduct it on VCTK~\cite{VCTK} and test-clean subset of LibriSpeech. For VCTK, we randomly select data from 8 speakers and 2911 utterances for the test.
For LibriSpeech, we use the test-clean subset, 2620 utterances for the test. All audio samples are downsampled to 16 kHz.
\textbf{2) Disentanglement Ability}. we evaluate it based on the any-to-any voice conversion benchmark. We randomly select 200 utterances from LibriSpeech test-clean subset as source speech and 6 speakers from VCTK as the target speaker. All models are evaluated in LibriSpeech Test-clean to VCTK scenarios.

\noindent
\textbf{Subjective Evaluation.} 
We follow the established MUSHRA methodology~\cite{mushra} to evaluate the subjective quality of our baselines and FreeCodec-v2.
A group of fifteen listeners participate in the subjective tests.
Sixteen utterances are randomly selected from our test sets for evaluation.
In addition, we also adopt the Speex \cite{valin2016speex} at 4 kbps as our low anchor.

\noindent
\textbf{Objective Evaluation.} For objective evaluation of reconstruction, we employ the automatic Mean Opinion Score prediction system (UTMOS) \cite{saeki2022utmos}, and the short-time objective intelligibility (STOI)~\cite{2010stoi}, and the WARP-Q \cite{jassim2021warp}, and the Speaker Embedding Cosine Similarity (SECS)\footnote{\url{https://github.com/resemble-ai/Resemblyzer}} to evaluate the overall speech quality. 
In addition, we use Word error rate (WER), character error rate (CER), and F0-PCC for the objective evaluation of voice conversion. 
Among them, WER and CER between source and converted speech are calculated by an ASR model\footnote{\url{https://huggingface.co/openai/whisper-large}}.
F0-PCC is the Pearson correlation coefficient used to evaluate \textit{${f}_{0}$} consistency between source and converted speech. 
It can be used to evaluate the preservation of prosody information apart from speaker and content information.

\begin{table}[h]
\centering
\vspace{-0.5cm}
\caption{The Subjective evaluation of reconstruction quality under unseen speakers from VCTK and LibriSpeech Test-Clean corpus with the 95\% confidence interval for each score.}
\vspace{-0.2cm}
\label{tab:mushra}
\setlength{\tabcolsep}{0.3mm}{
\begin{tabular}{llccc}
\toprule
\multirow{2}{*}{Method} & \multirow{2}{*}{Bitrate} & Supervised & Speaker  & MUSHRA     \\
                        &                          & Data       & Decouple & Score      \\ \midrule
Target                  & -                        & -          & -        & 95.09$\pm$0.44 \\
Speex$^{\scalebox{0.6}{$\clubsuit$}}$                 & 4kbps                  & No         & No       & 24.03$\pm$2.15 \\
SpeechTokenizer$^{\scalebox{0.6}{$\clubsuit$}}$         & 3 kbps                   & No         & No       & 82.0$\pm$1.44  \\
FACodec$^{\scalebox{0.6}{$\clubsuit$}}$                 & 2.4 kbps                 & Yes        & Explicit & 80.44$\pm$1.29 \\
SemantiCodec$^{\scalebox{0.6}{$\clubsuit$}}$            & 1.3 kbps                 & No         & No       & 73.44$\pm$1.3  \\
TiCodec$^{\scalebox{0.6}{$\blacklozenge$}}$                 & 1 kbps                   & No         & Implicit & 83.0$\pm$1.10 \\
DAC$^{\scalebox{0.6}{$\blacklozenge$}}$                     & 1 kbps                   & No         & No       & \underline{85.5$\pm$1.19}  \\
Wavtokenizer$^{\scalebox{0.6}{$\clubsuit$}}$            & 0.9kbps                  & No         & No       & 78.56$\pm$2.54 \\
TiCodec$^{\scalebox{0.6}{$\blacklozenge$}}$                 & 0.5 kbps                 & No         & Implicit & 73.06$\pm$2.50 \\
DAC$^{\scalebox{0.6}{$\blacklozenge$}}$                     & 0.5 kbps                 & No         & No       & 76.56$\pm$2.35 \\
FreeCodec-v2            & 0.45 kbps                & No         & Explicit & \textbf{87.44$\pm$0.88} \\ \bottomrule
\end{tabular}%
}
\end{table}

\begin{table}[h]
\vspace{-0.3cm}
\centering
\caption[short]{The objective evaluation of disentanglement ability. For WER and CER, the smaller the better. F0-PCC ranges from -1 to 1 and the higher
the better. }
\vspace{-0.2cm}
 \setlength{\tabcolsep}{0.48mm}{%
\begin{tabular}{lllccc}
\toprule
\multirow{2}{*}{Method} & \multirow{2}{*}{Bandwidth$\downarrow$} & \multicolumn{4}{c}{Test-Clean to VCTK}                                          \\
           &          & WER$\downarrow$ & CER$\downarrow$ & F0 PCC$\uparrow$ & SECS$\uparrow$ \\ \midrule
FreeCodec-v3            & \textbf{0.45 kbps}                & \underline{8.37} & 6.14 & 0.702 & \textbf{0.847} \\
TiCodec$^{\scalebox{0.6}{$\blacklozenge$}}$    & 0.5 kbps & 35.74   & 25.19  & 0.680   & 0.656          \\
TiCodec$^{\scalebox{0.6}{$\blacklozenge$}}$    & 1 kbps & 8.83  & \underline{6.08}   &\underline{0.752}   & 0.607         \\
FACodec$^{\scalebox{0.6}{$\clubsuit$}}$    & 2.4 kbps & \textbf{2.83}   & \textbf{2.57}   & \textbf{0.755}   & 0.553          \\ \hline
YourTTS$^{\scalebox{0.6}{$\clubsuit$}}$    & -        & 9.20            & 6.92            & 0.682            & 0.815          \\
Wav2vec-vc$^{\scalebox{0.6}{$\clubsuit$}}$ & -        & 13.23           & 9.20            & -0.037           & 0.826          \\
VQMIVC$^{\scalebox{0.6}{$\clubsuit$}}$     & -        & 56.58           & 39.21           & 0.611            & 0.650          \\ \bottomrule
\end{tabular}%
\label{vc}
}
\end{table}

\section{Results}
\subsection{Reconstruction Quality}

Table \ref{speech} summarizes the results of objective reconstruction experiments. 
FreeCodec-v1 performs best or second-best in almost all objective metrics in test sets. Especially in out-of-domain environments, our proposed method achieves superior reconstruction performance using only approximately 57 tokens per second than existing methods, such as Wavtokenizer at 0.9 kbps, and SemantiCodec at 1.3 kbps. Even compared to FACodec at 2.4 kbps and SpeechTokenizer at 3 kbps, FreeCodec-v1 gets comparable performance at STOI and speaker similarity. Compared to FreeCodec-v2, FreeCodec-v1 is better especially in speaker similarity. It shows that the continuous global representation is more effective in reconstruction scenarios(e.g., zero-shot TTS).

Although STOI and SECS are slightly lower than DAC at 1 kbps, FreeCodec-v2 has better objective speech quality, according to UTMOS and WARP-Q. 
The same result can also be concluded in subjective evaluation, as illustrated in Table \ref{tab:mushra}. As illustrated in subjective results, we can observe that FreeCodec-v2 at 0.45 kbps significantly outperforms both FACodec at 2.4 kbps and SpeechTokenizer at 3 kbps. Additionally, compared to TiCodec at 1 kbps and DAC at 1 kbps, FreeCodec-v2 gets higher scores with the same experimental configuration. The results of the subjective evaluation show the absolute advantages of FreeCodec-v2 in reconstruction compared to existing methods. 

Furthermore, as shown in Table\ref{speech}, we also conducted an ablation study to validate the explicit effect of content loss in the content encoder.
It can be observed that removing the content loss causes the performance drop in all objective metrics, especially the UTMOS and STOI.

\begin{figure}[h]
\centering
\vspace{-0.3cm}
\begin{minipage}[b]{0.6\linewidth}
    {\subfigure[${Z}_{s}$ ]{\includegraphics[width=1\linewidth,height=0.8\linewidth]{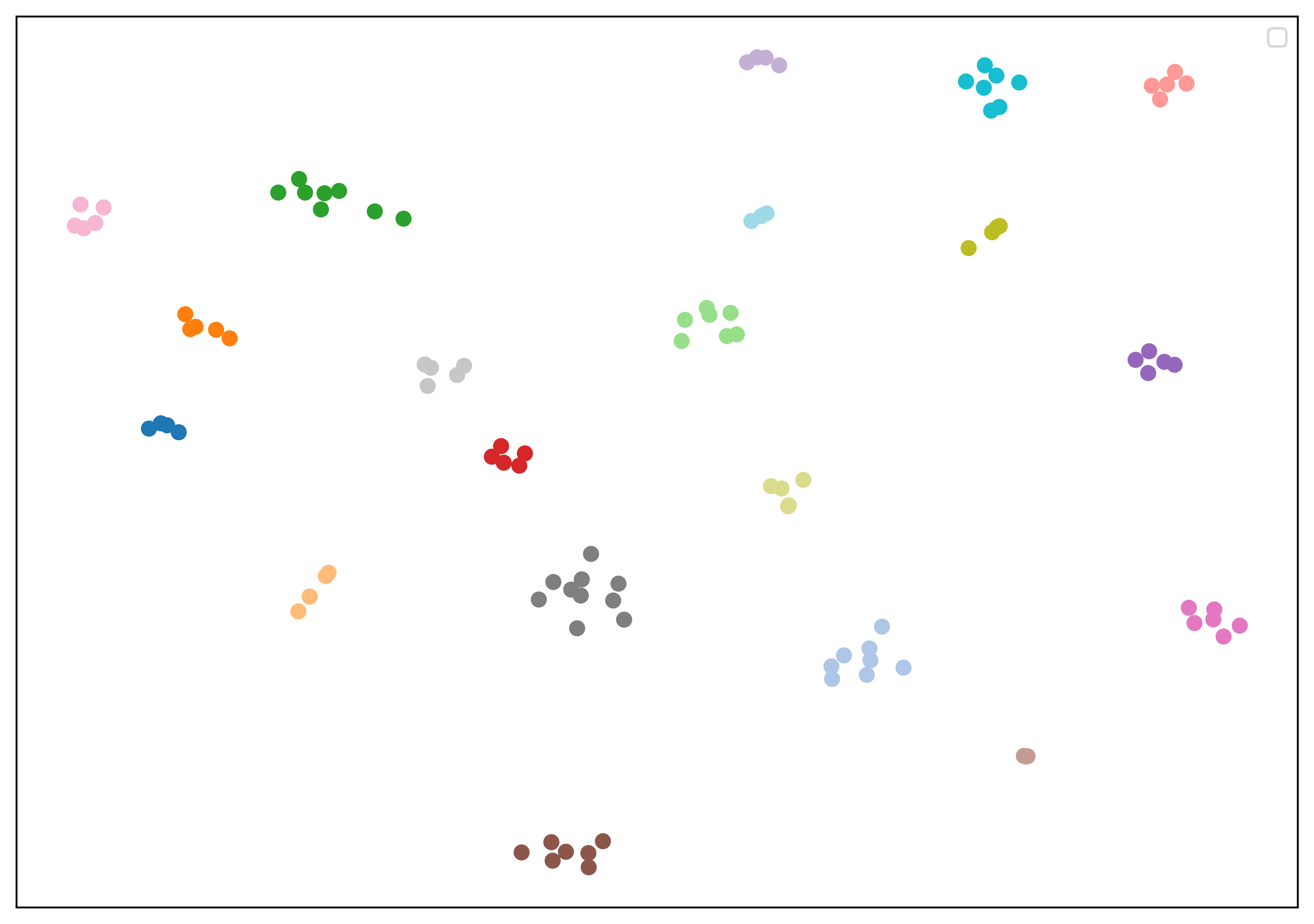}}}
\end{minipage}
\begin{minipage}[b]{0.28\linewidth}
    \subfigcapskip=-4.5pt
    \subfigbottomskip=0.1pt
    \subfigure[${Z}_{c}$ ]
    {\includegraphics[width=1.05\linewidth,height=0.82\linewidth]{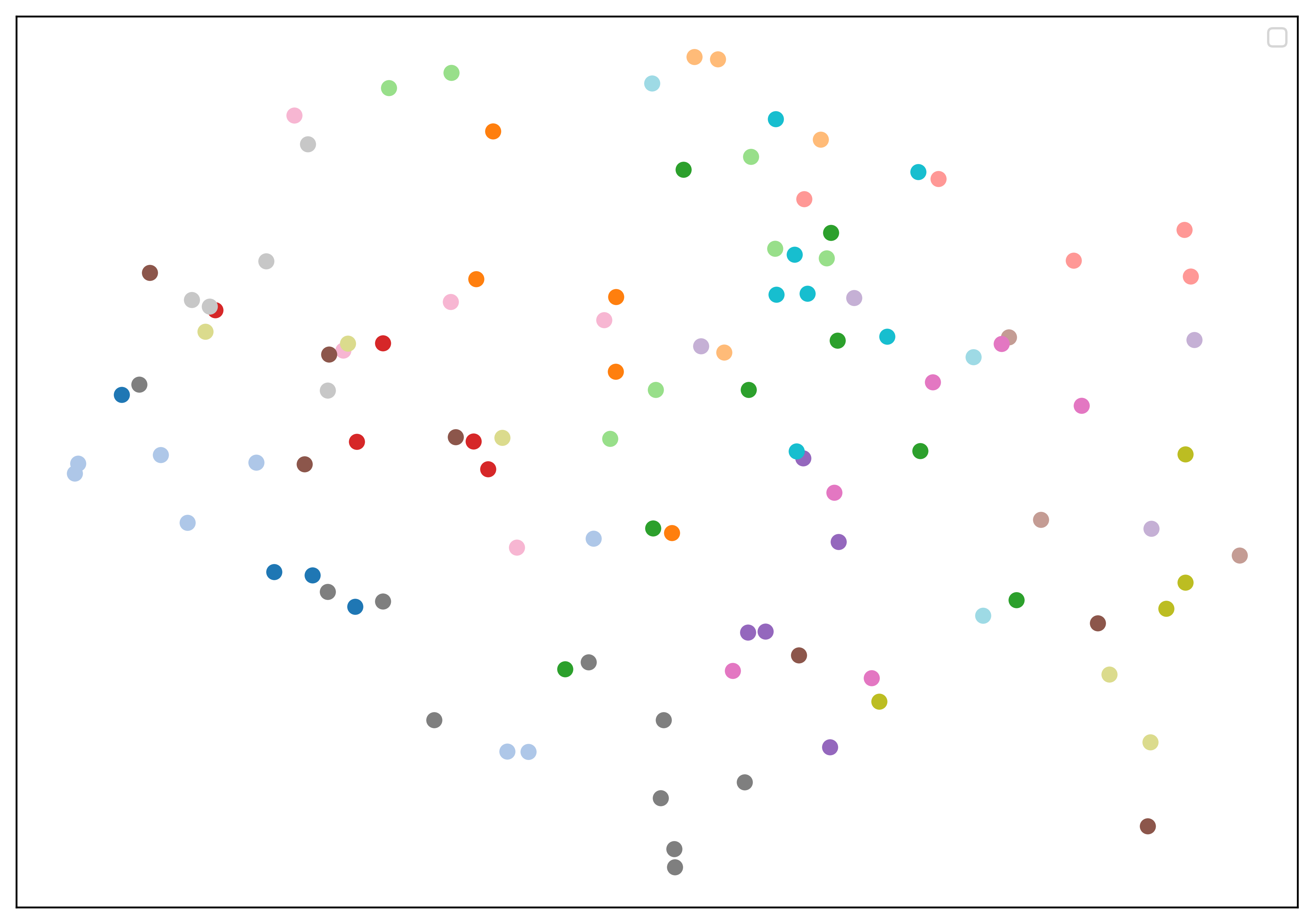}}
    \subfigure[ ${Z}_{p}$ ]
    {\includegraphics[width=1.05\linewidth,height=0.82\linewidth]{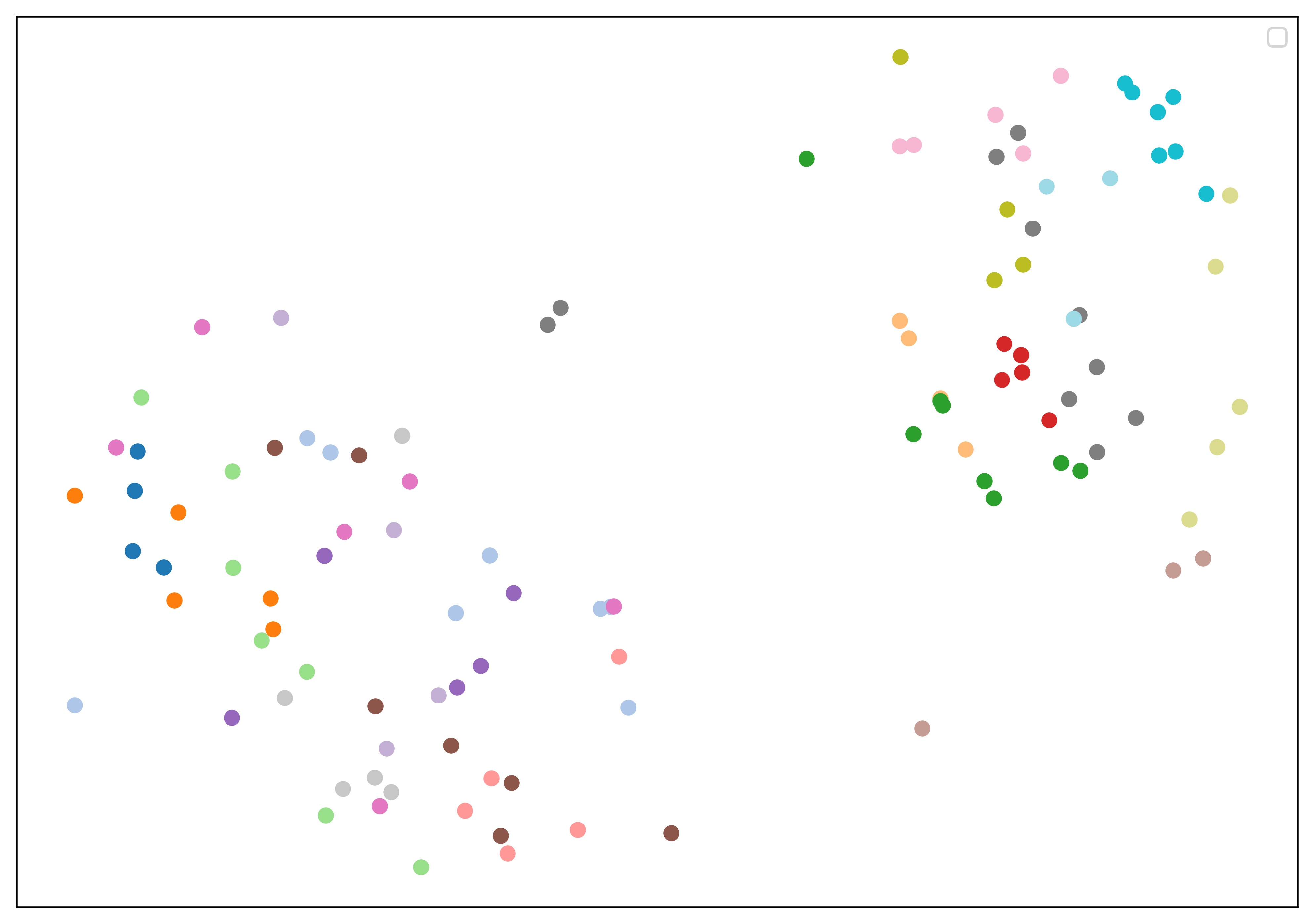}}
\end{minipage}
\vspace{-0.2cm}
\caption{(a)(b)(c) is the t-SNE visualization of speaker representations ${Z}_{s}$, content representations ${Z}_{c}$ and prosody representations ${Z}_{p}$ after VQ. Different colors represent different speakers.}
\vspace{-0.6cm}
\label{fig:tsne}
\end{figure}

\subsection{Disentanglement ability}

In this section, we describe the disentanglement ability on the voice conversion experiments. 
FreeCodec-v3 achieves voice conversion by using the speaker information from the target speech. 
As shown in Table \ref{vc}, FreeCodec-v3 at 0.45 kbps achieves the best speaker similarity in unseen speaker scenarios than all baseline models, especially supervised FACodec at 2.4 kbps and text-based models. The results show that our proposed method decouples the speaker information well under unseen-speaker scenarios in a self-supervised manner. 

Compared to TiCodec at 0.5 kbps and 1 kbps, FreeCodec-v3 exhibits lower WER and higher speaker similarity.  This indicates that our method preserves content information at ultra-low bitrates while achieving superior disentanglement compared to methods based solely on implicit bottlenecks. Additionally, FreeCodec-v3 shows comparable performance in F0 PCC, suggesting that our method effectively maintains the prosody of the source speech, achieving a better balance between prosody and speaker information.

We randomly selected 20 unseen speakers from the test sets and generated t-SNE visualizations to examine the distributions of speaker, content, and prosody representations. For content and prosody features, we collapsed the frame-level representations of each sample into a single vector through mean pooling. As shown in Fig. \ref{fig:tsne}, the speaker embeddings exhibit clear clustering patterns corresponding to individual speakers. In contrast, content representations appear distributed without discernible patterns across different speakers, while prosody features demonstrate moderate clustering tendencies for certain speakers. These observations collectively indicate that prosodic information maintains partial speaker-specific characteristics while remaining distinct from both speaker and content.


\vspace{-0.2cm}
\section{Conclusion}

In this paper, we propose a self-supervised disentanglement speech codec that factorizes speech into its intrinsic attributes. We demonstrate that this framework can be applied to both reconstruction and disentanglement tasks using different training strategies. Compared to existing methods, our approach utilizes fewer tokens and lower bandwidth while achieving high-quality reconstruction and superior disentanglement relative to supervised methods. 
Our experiments show a significant improvement over existing methods that use more than 2x bitrate, highlighting the effectiveness of our approach in reconstruction quality and disentanglement ability.

\section{Acknowledgement}
This paper is supported by the National Nature Science Foundation
of China (No. 62471343).

\bibliographystyle{IEEEtran}

\end{document}